\documentclass[fleqn,10pt]{wlscirep}
\usepackage[utf8]{inputenc}
\usepackage[T1]{fontenc}
\usepackage{graphicx}
\usepackage{amssymb}
\usepackage{color}
\usepackage{bbm}
\usepackage{braket,comment}
\usepackage{hyperref}

\newcommand{\ketbra}[2]{\ket{#1} \bra{#2}}
\newcommand{\e}{\text{e}}

\title{Open-loop quantum control of small-size networks for high-order cumulants and cross-correlations sensing}

\author[1]{Antonio D'Arrigo}
\author[2]{Giulia Piccitto}
\author[1,3,4]{Giuseppe Falci}
\author[1,3,4, *]{Elisabetta Paladino}
\affil[1]{Dipartimento di Fisica e Astronomia "Ettore Majorana", Universit\`a di Catania, Via Santa Sofia 64, 95123 Catania, Italy}
\affil[2]{Dipartimento di Matematica e Informatica, Universit\`a di Catania, Viale Andrea Doria, 95125 Catania, Italy}
\affil[3]{CNR-IMM, Catania (University unit), Consiglio Nazionale delle Ricerche, Via Santa Sofia 64, 95123 Catania, Italy}
\affil[4]{Istituto Nazionale di Fisica Nucleare, Sezione di Catania, Via Santa Sofia 64, 95123 Catania, Italy}
\affil[*]{elisabetta.paladino@dfa.unict.it}

\begin{abstract}
Quantum control techniques represent one of the most efficient tools to attain high-fidelity quantum operations and a convenient approach
for quantum sensing and quantum noise spectroscopy. In this work, we investigate dynamical decoupling while processing an entangling two-qubit gate based on an Ising-$xx$ interaction, each qubit being affected by pure dephasing classical correlated $1/f$-noises. To evaluate the gate error, we used the Magnus expansion introducing generalized filter functions that describe decoupling while processing and 
allow us to derive an approximate analytic expression as a hierarchy of nested integrals of noise cumulants. The error is separated 
in contributions of Gaussian and non-Gaussian noise, the corresponding generalized filter functions being calculated up to the fourth order.  By exploiting the properties of selected pulse sequences, we show that it is possible to extract the second-order statistics (spectrum and cross-spectrum) and to highlight non-Gaussian features contained in the fourth-order cumulant. We discuss the applicability of these results to state-of-the-art small networks based on solid-state platforms.
\end{abstract}

\begin{document}
\flushbottom
\maketitle
\thispagestyle{empty}

\section{Introduction}
\label{intro}

In the current generation of solid-state devices for quantum technologies~\cite{preskill_quantum_2018}, environmental noise sets the accuracy limits of quantum gates~\cite{Kjaergaard_2020}. Despite the tremendous progress in the last two decades~\cite{Gambetta2017,Siddiqi2021,FALCI20241003}, 
material-inherent noise sources still represent a 
problem making unreliable even moderate-size quantum circuits.
Quantum control techniques~\cite{wiseman2010book}  
represent one of the most efficient tools to attain high-fidelity quantum operations fulfilling given time and power constraints. Their primary goal is to maintain noise-induced errors below a fault-tolerance threshold required for the efficient implementation of quantum error correction. Dynamical decoupling (DD)~\cite{viola1998PRA,viola1999PRL,vitali1997PRL} is a form of open-loop quantum control whose efficiency has been repeatedly validated in experiments using a variety of platforms~\cite{Biercuk2009, biercuk2009Nature,deLange2010,Bluhm2011,Medford2012}.
The effect of DD can be seen as a noise filtering process~\cite{biercuk2011JPB} mathematically expressed in terms of (generalized) filter functions (FFs)~\cite{Green2012,Green2013,Soare2014,Frey2017}. 

From a different perspective, dynamical control can be turned into a tool for quantum sensing (QS) and quantum noise spectroscopy~\cite{Degen2017} whereby properly designed pulsed ~\cite{Yuge2011PRL, Alvarez2011,Norris2016,Faoro2004PRL, falci2004PRA,Zwick2016,szczykulska2016,nazarov2012,ernst1990} 
or continuous-control protocols~\cite{Yan2013,Frey2017,Norris2018,Frey2020,vonLupke2020,ka:216-vepsalainen-photonics-squtrit,ka:19-falci-scirep-usc} allow inferring microscopic information, as noise power spectra. This provides complete statistical information on Gaussian processes whereas the characterization of non-Gaussian fluctuations requires estimating higher-order correlation functions, or polyspectra in Fourier space~\cite{2016Szank,Ramon2019}. Discriminating this type of information is of paramount relevance in state-of-the-art devices~\cite{vonLupke2020,Rower2023,Trappen2023} where evidence of microscopic two-level systems either coherently coupled to the quantum circuit or incoherently evolving like random telegraph noise (RTN) processes,  
has been demonstrated both in spectroscopy and in time-domain measurements~\cite{PaladinoRMP}. 
RTN is the archetypical non-Gaussian process and higher-order spectral estimation using a qubit probe under pulsed control
\cite{Norris2016,Ramon2019,Sung2019} or via a frame-based control-adapted FF formalism~\cite{Dong2023} have been recently demonstrated.  Correlated Gaussian processes~\cite{PazSilva2017,2019Krzywda} and RTN~\cite{2016Szank} inducing pure dephasing have been investigated via multipulse quantum noise spectroscopy protocols. Collections of RTNs with proper distribution of switching rates are a common model of noise with power spectrum behaving as $1/f^\alpha$ where 
$f$ is the frequency and $\alpha \sim 1$~\cite{weissman1988RMP}.
Inherently microscopic noise sources with $1/f$-like spectral density and/or non-Gaussian characteristics~\cite{PaladinoRMP} represent one of the major problems for quantum state processing of state-of-the-art scalable solid-state qubits\cite{PaladinoRMP,FALCI20241003,Yoshihara2006,Kakuyanagi2007, Bialczak2007, Koch2007, Bylander2011,ka:20-pellegrino-commphys-1overfgraph,McCourt2023,ka:23-pellegrino-apl-secondspec}. Preservation of entanglement in the presence of RTN or $1/f$ noise via DD protocols has been also proposed~\cite{Wang2011,LoFranco_2013,LoFranco2014, DarrigoPhysSCr2015,Pokharel2018,Yan2022}.

DD of local and spatiotemporal correlated noise sources with $1/f$ or RTN spectrum in an entangling gate is a critical step to achieving high-fidelity two-qubit gates. 
This issue received so far less attention, despite recent experiments revealing spatial noise correlations~\cite{vonLupke2020,Yoneda2023}.
DD of a two-transmon gate with a noisy tunable coupler has been recently investigated~\cite{McCourt2023}, pointing out the role of $1/f$ flux noise in the coupler and observing non-Gaussian signatures analogous to those investigated in single qubit gates~\cite{paladino2002PRL,galperin2006PRL}.  

In this work, we consider an entangling two-qubit gate based on an Ising-$xx$ interaction 
with strength $\omega_c$. Each qubit is affected by local pure dephasing classical noises with power spectrum $S(f) = A/f$ in the range of frequencies $f \in [f_m, f_M]$ with some degree of correlation~\cite{PaladinoRMP} quantified by a non-vanishing cross-spectrum $S_c(f)$ between the random forces. We consider processes characterized by $S_c(f)= \mu \, A/f$, 
where the parameter $\mu \in [0,1]$ quantifies the strength of the correlations ~\cite{Zorin1996,vonLupke2020,McCourt2023}. 

We study DD  protocols implemented by sequences of instantaneous pulses acting on each qubit locally and simultaneously designed in a way not to alter the capability of the gate to generate entanglement at a time $t_e= \pi/(2 \omega_c)$. We evaluate the gate error, both in the time and in the frequency domain, using a Magnus expansion technique. For local longitudinal noise, 
the evolution is exactly mapped to two-level problems with {\em transverse} coupling to classical noise. By following an approach inspired by~\cite{Green2012,Green2013}, we derive an approximate analytic expression for the error as a hierarchy of nested integrals of noise cumulants and FFs.  
Depending on the DD sequence and the statistical properties of the noise, the gate error is dominated by contributions of cumulants of different order.  Up to the fourth order, we can separate the error in contributions due to Gaussian and non-Gaussian components identifying the corresponding FFs. The different scaling of these terms with the correlation parameter $\mu$ allows the characterization of the noise statistics and cross-correlations. By exploiting the filtering properties of the DD sequences considered, we show that it is possible to extract the second-order statistical properties (spectrum and cross-spectrum) and to highlight non-Gaussian features by the fourth-order cumulant.

\section{Results}\label{sec:results}
\subsection{The protocol}
We start considering a system of two coupled identical qubits in the presence of classical noise described by the Hamiltonian ${\cal H}(t) = \mathcal{H}_0 +  \delta \mathcal{H}(t)$ where
(units of $\hbar=1$ are chosen)
\begin{equation}
\mathcal{H}_0 = 
 -\frac{\Omega}{2} \, \sigma_{1z} \otimes \mathbbm{1}_{2}
 -\frac{\Omega}{2}  \, \mathbbm{1}_{1} \otimes \sigma_{2z}
+ \frac{\omega_c}{2} \, \sigma_{1x} \otimes \sigma_{2x},
\quad 
\delta \mathcal{H}(t) = 
-\frac{z_{1}(t)}{2} \, \sigma_{1z} \otimes \mathbbm{1}_{2} - \frac{z_{2}(t)}{2} \mathbbm{1}_{1} \otimes  \sigma_{2z},
\label{eq:H}
\end{equation}
where $\sigma_{\alpha x}$ and $\sigma_{\alpha z}$ are the Pauli operators acting on the qubit $\alpha$, 
being the logic basis such that $\sigma_{\alpha z} \ket{\pm}_{\alpha} = \mp\ket{\pm}_\alpha$. 
When the qubits natural frequencies $\Omega$ are 
much larger than the coupling strength $\omega_c$, the evolution for a time $t_e = \pi/2\omega_c$ under $\mathcal{H}_0$ implements a $\sqrt{i-\text{SWAP}}$ two-qubit gate which has been demonstrated on different hardware platforms~\cite{chatterje2021,Kjaergaard_2020,Chen2023arXiv,Zhang2023arXiv}. For the sake of presentation, we focus on the effects of local classical noise longitudinally coupled to each qubit i.e. noise enters  $\delta{\cal H}(t)$ with terms commuting with the individual qubit Hamiltonian and discuss this choice later (see \S~\ref{sec:discussions}). Noise is modelled by two stochastic processes $z_\alpha(t)$ assumed to be of 4-th order stationary and characterized by their power spectra $S_{z_\alpha}(\omega)$ and fourth-order cumulants. Correlations of noises on different qubits are quantified 
by the cross-spectrum $S_{z_1 z_2}(\omega)$ (see supplemental \S~\ref{supp:stochastic}). 

Control is operated by applying simultaneously to both qubits a sequence made of an even number of $\pi-$pulses around the $y-$axis of the Bloch sphere as described by the Hamiltonian ${\cal H}_C(t)$ in Eq.~\eqref{eq:control-sequence}.
This protocol is designed to dynamically decouple the system from the noisy environment while executing a two-qubit gate. Indeed, in the asymptotic limit of a large number of pulses, the sequence averages out the diagonal terms of the Hamiltonian while keeping the coupling term $\sigma_{1x} \otimes \sigma_{2x}$. 
The error in the gate operation is quantified by 
\begin{equation}
\varepsilon\,=\,1-\bra{\psi_e}\rho(t_e)\ket{\psi_e},
\label{eq:gate-fidelity-error}
\end{equation}
where $\ket{\psi_e}$ is the target state of the ideal operation and $\rho(t_e)$ is the actual state at $t=t_e$ obtained from the evolution under the action of the controlled noisy Hamiltonian ${\cal H}(t)+{\cal H}_C(t)$. The gate infidelity is the maximal error with respect to the initial state $\ket{\psi_0}$. 

Under the action of $\cal H$, the system evolves in two invariant subspaces  (see suppl. \ref{supp:ham}). We focus on the dynamics in the single-excitation subspace $W= \mathrm{span}\{\ket{+-},\ket{-+}\}$. In the basis of the Bell states $\ket{\beta} = \big[\ket{+-}+ (-1)^\beta \ket{-+}\big]/\sqrt{2}$ for $\beta=1,2$ (see tab.~\ref{tab:eigensystem0}), the projected Hamiltonian reads
\begin{equation}
 {\cal H}_{W}(t) :=  P_W \, {\cal H} P_W  = -\, {\frac{\omega_c}{2}}\, \tau_z  - \frac{z_1(t) - z_2(t)}{2} \,\tau_x \;,
\label{eq:equivalent-2Dim-noisyHamiltonian}
\end{equation}
where $P_W$ are projection operators and $\tau$'s are Pauli matrices, $\tau_z = \ketbra{1}{1}- \ketbra{2}{2}$ and $\tau_x = \ketbra{1}{2} + \ketbra{2}{1}$. 
Therefore the effective dynamics in the $ W$ subspace is governed by a two-state Hamiltonian. The ideal gate generated by $\sigma_{1x} \otimes \sigma_{1x}$ is projected in a two-level unitary of the $W$-subspace which operates as a non-trivial quantum gate. The effective noise enters via the stochastic process $\zeta(t)=z_1(t)-z_2(t)$ which couples by an operator {\em transverse} to the 
projected ideal Hamiltonian $P_W \, {\cal H}_0 P_W$. 

In particular, we study the generation of a maximally entangled state obtained from the initial factorized state $\ket{\psi_0} = \ket{+-}$ by evolving the system in the absence of noise for a time $t_e$
\begin{equation}
\label{eq:entanglement-generation}
\ket{\psi}_e = e^{-i \mathcal{H}_0 t_e} \ket{+-} = \frac{\ket{+-} - i \ket{-+}}{\sqrt 2}.
\end{equation}
In the following, we focus on the error $\varepsilon$ for this operation.
This quantity will be used as the output of a QS protocol and it also provides an indicator of the gate fidelity in the $W$-subspace since the chosen $\ket{\psi_0}$ approximately maximizes the gate error for $\zeta \ll \omega_c$.

\subsection{Gate error under dynamical control}
We notice that the dynamics under the $y-y$ pulse sequence preserve the invariant subspaces of $\cal H$. Therefore, under DD control, the gate error for the operation Eq.~\eqref{eq:entanglement-generation} 
does not contain contributions due to leakage from the $W$ subspace.
One of the key results of this work is the following formula
expressing the gate error $\varepsilon$  as an expansion in the time-correlations of the noise truncated at the fourth-order, with the analytic form for the FFs $F_i(\omega,\omega_c,t_e,2n)$ reported in App. \ref{supp:filter}
\begin{equation}
\begin{aligned}
	\varepsilon\,= \varepsilon^{[2]} + \varepsilon^{[4]}_g + \varepsilon_{ng}^{[4]} = \,&\int_{-\infty}^{+\infty}\frac{d\omega}{2\pi}\,S_\zeta(\omega)
F_{1}(\omega,\omega_c,t_e,2n)\, \\
	+ \,&\int_{-\infty}^{+\infty}\frac{d\omega_1}{2\pi}\,S_\zeta(\omega_1)
\int_{-\infty}^{+\infty}\frac{d\omega_2}{2\pi}\,S_\zeta(\omega_2)\,
F_{2,g}(\vec \omega_2,\omega_c,t_e,2n)\,\\
+\,&\int_{-\infty}^{+\infty}\frac{d\omega_1}{2\pi}\,
\int_{-\infty}^{+\infty}\frac{d\omega_2}{2\pi}\,\int_{-\infty}^{+\infty}\frac{d\omega_3}{2\pi}
S_{\zeta 3}(\omega_1,\omega_2,\omega_3)\,
F_{2,ng}(\vec\omega_3,\omega_c,t_e,2n) \, .
\end{aligned}
\label{eq:error_formula}
\end{equation}
The second-order $\varepsilon^{[2]}$ depends on the power spectrum $S_\zeta (\omega)$ of the noise $\zeta(t)$. The latter is the sum of the power spectra of each physical process $z_\alpha(t)$ and of their cross-correlation (see supplemental \S~\ref{supp:stochastic}), $S_\zeta(\omega)= S_{z_1}(\omega) +S_{z_2}(\omega) - 2 S_{z_1 z_2} (\omega)$.
The fourth-order $\varepsilon^{[4]}$ can be written in general as the sum of a Gaussian ($\varepsilon^{[4]}_g$, second line) and a non-Gaussian ($\epsilon_{ng}^{[4]}$, third line) contribution. This latter depends on the trispectrum $S_{\zeta 3}(\omega_1,\omega_2,\omega_3)\,$ which is the Fourier transform of the (stationary) fourth-order cumulant of $\zeta(t)$. 

For a fixed duration $t_e$ of the gate operation, we analyze the dynamics under ${\cal H}_C(t)$ for three different sequences of $2n$ pulses, namely the periodic (P), the Carr-Purcell (CP) and the Uhrig (U) sequences (see \S~\ref{sec:methods} for details). Information on the pulse-sequence enters Eq.~\eqref{eq:error_formula} via the FFs $F_i(\omega,\omega_c,t_e,2n)$. Notice that our $F_i$s generalize the FFs used for standard DD and QS of longitudinal noise~\cite{Biercuk2009}. In our case a non-trivial gate operation is performed during the DD sequences, thus noise operators do not commute anymore with the Hamiltonian ${\cal H}_0$. The expression of the generalized FFs, which in our case explicitly depend on the coupling $\omega_c$, has been evaluated by exploiting the Magnus expansion of the evolution operator (see \S~\ref{supp:magnus}, \S~\ref{supp:filter}).
 
\begin{figure}[t!]
\centering
	\includegraphics[width=\textwidth]{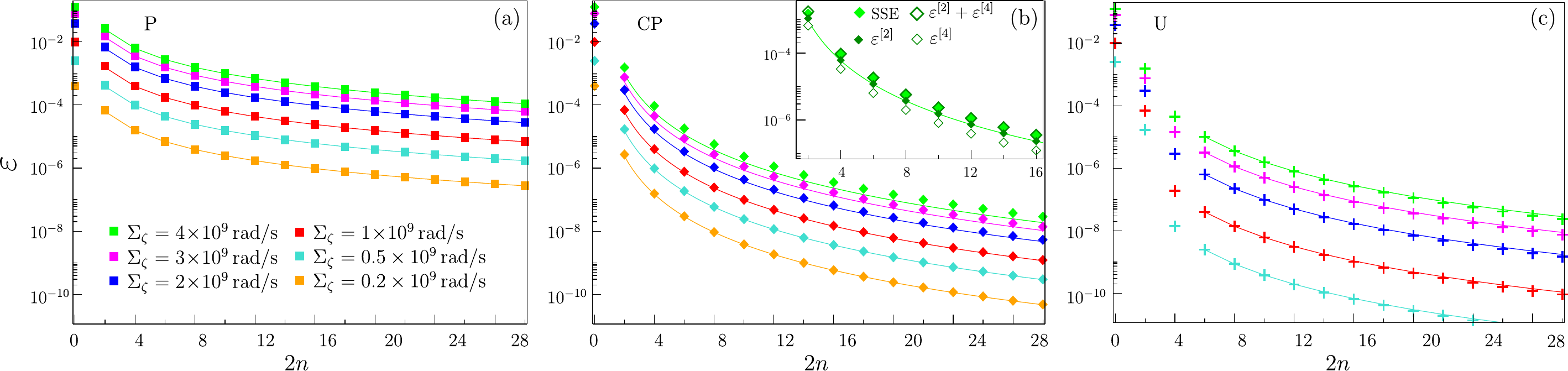}
	\caption{Gate error $\varepsilon$ for P (panel (a)), CP (panel (b)) and U (panel (c)) sequences as a function of the number of pulses $2n$ for fixed gate time $t_e=\pi/2\omega_c$, with  $\omega_c = 5 \cdot 10^9 \text{rad}/\text{sec}$ and different noise amplitudes $\Sigma_\zeta$. 
Symbols are data from the numerical solution of the SSE, the analytical expressions~\eqref{eq:static_error_P_C} for P and CP sequences, and  Eq.~\eqref{eq:static_error_U} for U sequence are given by the continuous lines.
Inset of the panel (b): different contributions to $\varepsilon$ for the CP sequence for $\Sigma_\zeta=4\times 10^9$ and $\omega_c=10^9$ rad/s: light green filled diamonds are the numerical solution of the SSE, large dark green diamonds are the error $\epsilon$ Eq.~\eqref{eq:error_formula}. The second and the fourth-order contributions to the error correspond to the dark green filled diamonds 
and small dark green diamonds respectively. }
\label{fig:variSigma-quasistatic-noise}
\end{figure}

A first insight into the problem is gained by considering  Gaussian quasi-static noise with variance $\Sigma_\zeta^2$. In this limit, only frequencies much lower than $\omega_c$ enter Eq.~\eqref{eq:error_formula}, thus the power spectrum can be approximated by
$S_\zeta(\omega) = 2 \pi \Sigma^2_\zeta\delta(\omega)$.
In Fig.~\ref{fig:variSigma-quasistatic-noise} we show $\varepsilon$ for the pulse sequences under study and various $\Sigma_\zeta$. 
The symbols are the numerical solution of the stochastic Schr\"odinger equation (SSE). The filled lines in panel (a) and panel (b) are the following analytical expressions
\begin{equation}
    \varepsilon_{qs}^{(P)} \simeq \frac{\pi^2}{2^6}\left(\frac{\Sigma_\zeta}{\omega_c}\right)^2\frac{1}{n^2} \quad , \quad  
    \varepsilon_{qs}^{(CP)} \simeq \frac{\pi^4}{2^{12}}\left(\frac{\Sigma_\zeta}{\omega_c}\right)^2\frac{1}{n^4} \;,
	\label{eq:static_error_P_C}
    \end{equation}
derived from Eq.~\eqref{eq:error_formula} by substituting the power spectrum. We notice that there is an excellent agreement between numerics and analytical approximations. 
Both sequences suppress noise for increasing pulse rate and decreasing ratio ${\Sigma_\zeta}/{\omega_c}$, in agreement with the analogy between DD and the Zeno-effect~\cite{facchi_unification_2004}. Moreover, even though both errors scale quadratically with $\Sigma_\zeta$, the CP ($\propto 1/n^4$) produces, for increasing pulse rate, stronger error suppression than the P sequence ($\propto 1/n^2$). The numerical analysis for the CP sequence suggests that, for $\Sigma_\zeta > 10^9$ rad/s, the second and the fourth-order terms contribute $\varepsilon$ with comparable magnitude. 
This is shown in the inset of panel (b) where we plot the second-order and the fourth-order contribution, $\varepsilon_g$ and the solution of the SSE, for $\Sigma_\zeta = 4\times10^9$ rad/s.

The approximate analytic behaviour for the U sequences  
(filled lines in Fig.~\ref{fig:variSigma-quasistatic-noise}(c))
is derived by fitting the SSE numerical result. 
For $n > 4$ we found
\begin{equation}
    \varepsilon_{qs}^{(U)} \propto \left(\frac{\Sigma_\zeta}{\omega_c}\right)^4\frac{1}{n^{3.6}}.
	\label{eq:static_error_U}
\end{equation}
The dependence on $\Sigma_\zeta^4$ suggests that the Uhrig filter fully cancels the contribution of the second-order time-correlation function during processing in the presence of transversal effective noise.
\begin{figure}[t!]
\centering
\includegraphics[width=0.7\textwidth]{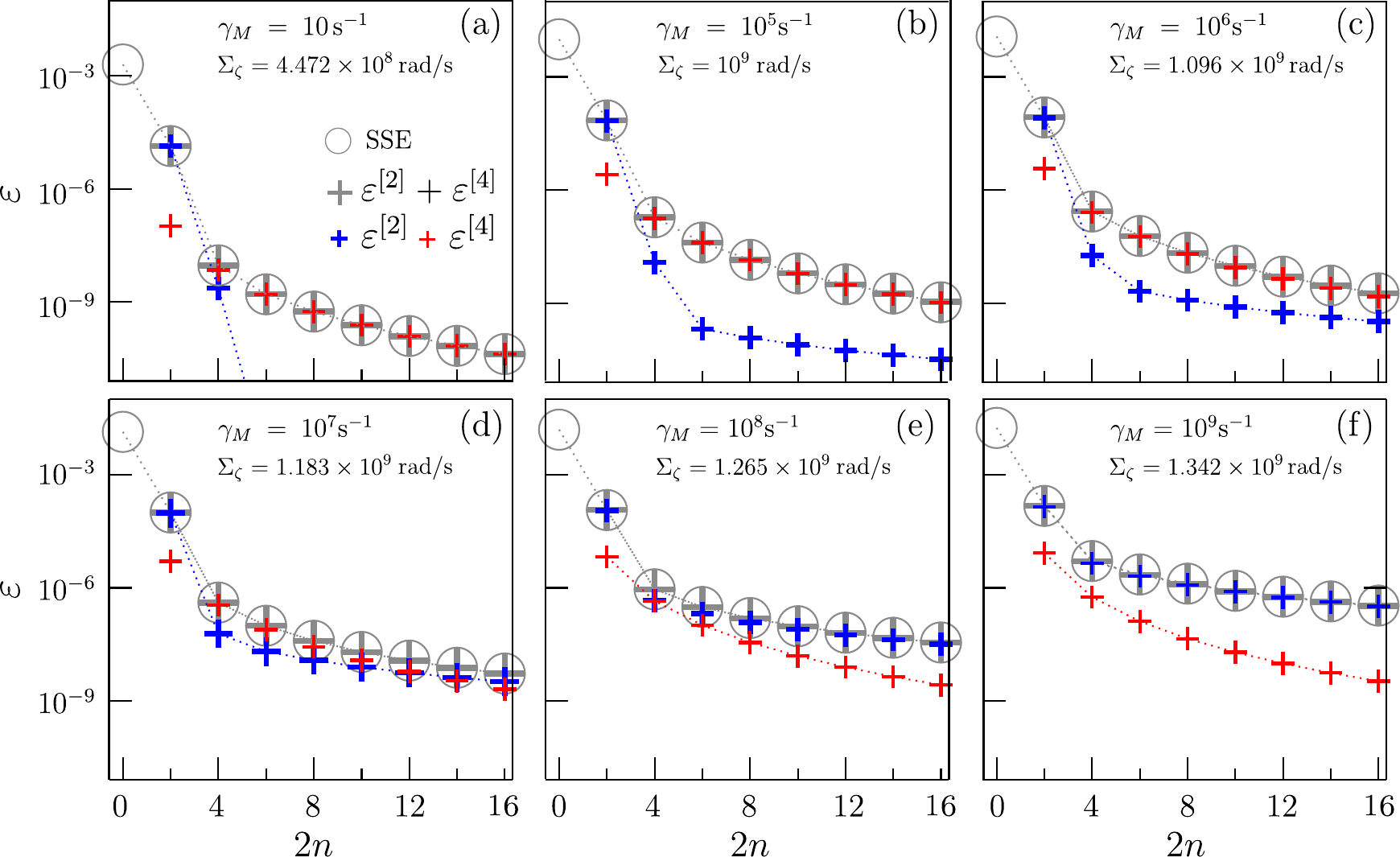}
	\caption{Gate error for the U sequence as a function of $2n$ for a Gaussian $1/f$ noise with fixed low-frequency cutoff $\gamma_m=1s^{-1}$ and for 
    $\omega_c = 5 \cdot 10^9$ rad/s.
    Different panels refer to different upper cut-off $\gamma_M$ and 
    $\Sigma_\zeta$ is chosen such as to have the same integrated power spectrum for any $\gamma_M$ (parameters in panel (b) correspond to typical values of charge noise in superconducting qubits). 
	In each panel, open circles are the solutions of the SSE equation, grey crosses give the error of Eq.~\eqref{eq:error_formula}, while blue and red crosses represent the 2nd and the 4th order Gaussian contributions, respectively. The Uhrig sequence practically cancels the 2nd order contribution for $\gamma \le 10^{6} s^{-1}$.}
\label{fig:figure:UDD-different-noise-cutoff}
\end{figure}

Remarkably, the same results can be derived for the case of long-time correlated noise. 
In Fig.~\ref{fig:figure:UDD-different-noise-cutoff} we report the gate error $\varepsilon$ in the presence of noise with spectrum
$S_\zeta(\omega)=A/\omega$ having fixed low-frequency cut-off $\gamma_m=1s^{-1}$ and varying the upper cut-off $\gamma_M$.  
Open circles are the solution of the SSE equation, grey crosses are obtained by Eq.~\eqref{eq:error_formula}, and blue and red crosses represent respectively the second and the fourth-order contribution in Eq.~\eqref{eq:error_formula}. The analytical approximation Eq.~\eqref{eq:error_formula} and the numerical SSE are in agreement also in this case. This result assesses the ability of the DD procedure to suppress noise at $\omega_c \neq 0$ and yields also in this case the Zeno effect scenario. For low-frequencies noise ($\gamma_M \le 10^6 \ \text{s}^{-1}$) the error is dominated by the fourth-order term, while the situation is reversed and the second-order contribution becomes dominant, as soon as high-frequency noise enters the game, i.e.  $\gamma \ge 10^{-8} s^{-1}$. This behaviour can be used for applications to QS.

\subsection{Filter functions and quantum sensing}

\begin{figure}[h!]
\centering
\includegraphics[width=0.9\textwidth]{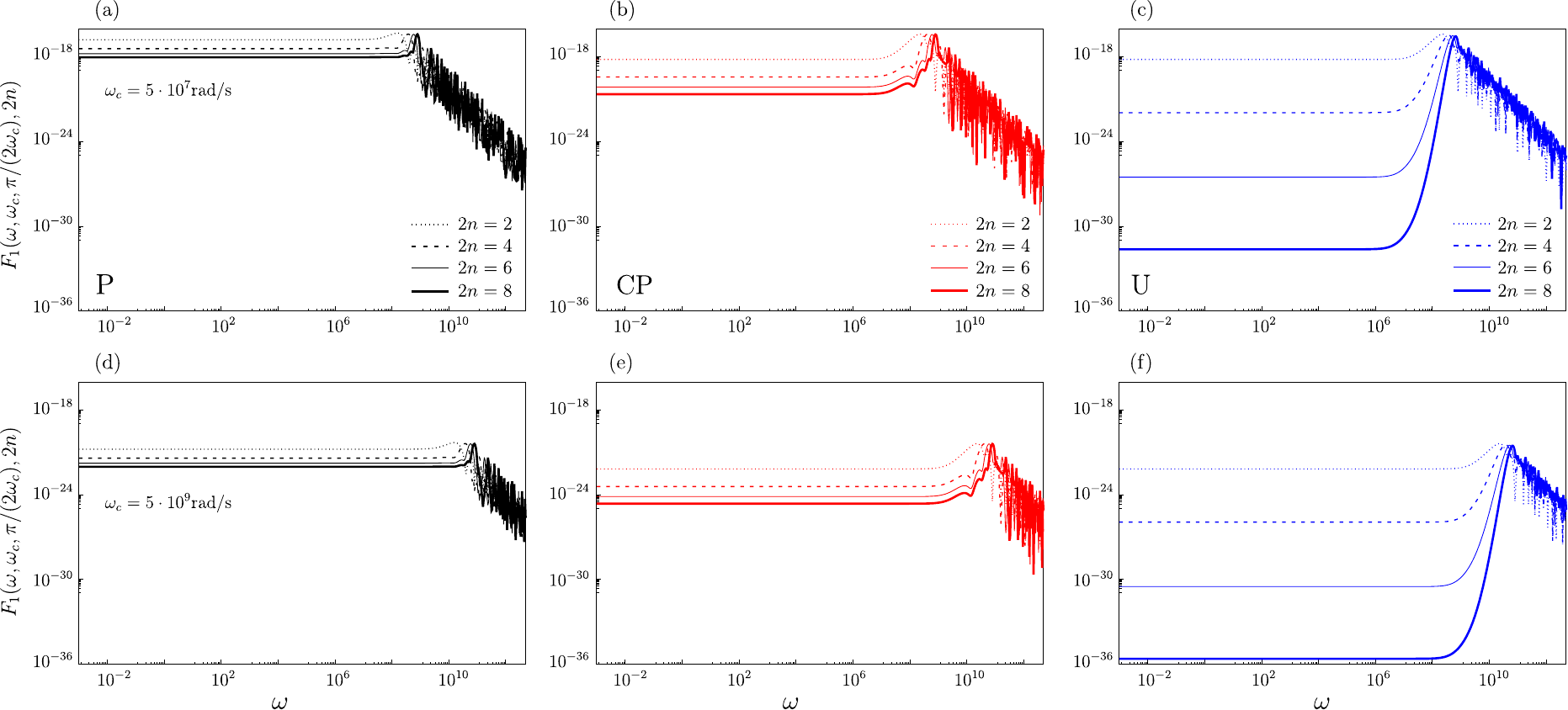}
	\caption{Plots of the second order FF $F_{1}^{}(\omega,\omega_c,t_e,2n)$, with $t_e=\pi/(2\omega_c)$, as a function of $\omega$.  We use $\omega_c=5\cdot 10^7$ rad/s in the top panels and $\omega_c=5\cdot 10^9$ rad/s for the bottom ones. Different colours refer to different sequences: P (black, panels (a) and (d)), CP (red, panels (b) and (e)) and U (blue, panels (c) and (f)). In each figure, the curves refer to (from top to bottom) $2n=2,\, 4,\,6,\, 8$. The U-filter suppresses low-frequency noise by several orders of magnitude already with a small number of pulses. }
\label{fig:2ndFilterFunctionplots-1}
\end{figure}

Standard FFs~\cite{biercuk2009PRL} can be designed to increase the protection of coherence from longitudinal noise using DD techniques~\cite{Biercuk2009}. 
Moreover, modulation of properly designed filters is a powerful tool for QS of noise~\cite{Degen2017}, in particular from the perspective of experimentally characterizing the longitudinal noise with single-qubit~\cite{Yan2012, Alvarez2011,Norris2016,Faoro2004PRL, falci2004PRA,Zwick2016}. 
In this section, we investigate the behaviour of our generalized FFs. The main result suggests that the pulse sequences can provide distinct signatures of the local dephasing bath as non-Gaussianity and spatial correlations for $1/f$-like noise.
In Fig.~\ref{fig:2ndFilterFunctionplots-1} we show $F_{1}(\omega,\omega_c,t_e,2n)$ for $t_e=\pi/(2\omega_c)$ appearing in Eq.~\eqref{eq:a1y}, for the three pulse sequences introduced previously.
Regardless of the sequence considered, the filter has a maximum at $\omega \sim \omega_c$, which becomes sharper for increasing the number of pulses $n$. 
For frequencies $\omega\ll \omega_c$ the filter has a plateau
whose magnitude decreases with increasing the number of applied pulses.
The ratio between the value at the peak frequency and the magnitude of the low plateau for $n \geq 3$ is moderate for the P generalized filter, but it may be very large for the U one. 

\begin{figure}[h!]
\centering
\includegraphics[width=0.6\textwidth]{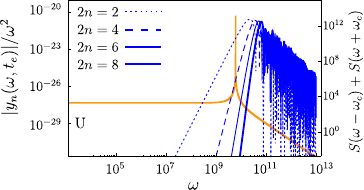}
\caption{The modified $1/f$ power spectrum $S(\omega-\omega_c)+S(\omega+\omega_c)$ (orange curve) in Eq.\eqref{eq:error-gate-FFA-2ndorder_bis} for  $\omega_c=5\cdot 10^9$ rad/s, 
$\gamma_{m}= 10^0$ s$^{-1}$, $\gamma_{M}= 10^6$ s$^{-1}$ and $\Sigma_\zeta= 4\cdot 10^8$ s$^{-1}$. It behaves as a narrow filter for 
the second-order $F_1(\omega,t_e,2n)$ 
weakening the impact of low-frequencies. For the Uhrig filter $F_1^U(\omega, t_e,2n)$ (blue curves)
and $t_e = \pi/(2 \omega_c)$, frequencies up to $\sim~ 2 \pi/t_e$ are 
suppressed already for $2n=6$.}
\label{fig:Urhig_filters}
\end{figure}

In particular, the U protocol efficiently filters out low-frequency noise in second order.
As a consequence, a relatively small number of pulses is enough to suppress the error $\varepsilon$ by several orders of magnitude.
To clarify this property we first observe that the behaviour of the generalized second-order FFs  $F_1(\omega,\omega_c,t_e, 2n)$ shown in Fig.~\ref{fig:2ndFilterFunctionplots-1}  is significantly different than that of the standard ones $F_1(\omega,t_e, 2n)$ for pure-dephasing noise, the blue line in
Fig.~\ref{fig:Urhig_filters}.
We can make a connection by rewriting the error $\varepsilon^{[2]}$ in Eq.~\eqref{eq:error_formula} in terms of the standard filters
\begin{equation}
\varepsilon^{[2]} \,=\,\frac{1}{16}\int_{-\infty}^{+\infty}\frac{d\omega}{2\pi} 
\Big[ S_{}(\omega-\omega_c) + S_{}(\omega+\omega_c)\Big] \, F_1(\omega,t_e,2n) \;.
\label{eq:error-gate-FFA-2ndorder_bis}
\end{equation}
The sum of the shifted spectra  $S_{}(\omega-\omega_c) + S_{}(\omega+\omega_c)$, shown in orange in Fig.(\ref{fig:Urhig_filters}),
has a sharp peak at $\omega_c$, while at smaller frequencies it exhibits a plateau whose value is orders of magnitude smaller than the original $1/f$ spectrum $S(\omega)$. Hence, it strongly weakens the impact of lower frequencies and
behaves as a sort of narrow filter at frequency $\omega_c$ for the $F_1(\omega,t_e,2n)$. 
Thus, by using the Uhrig sequence, we can leverage the properties of the standard Uhrig filter $F_1^U(\omega,t_e,2n) ={\big|y_n(\omega,t_e)\big|^2}/{\omega^2}$  (see sec. \ref{sec:filterfunction})
to suppress frequencies up to $\sim 2 \pi/t_e = 4 \omega_c$ very quickly as the number of pulses increases. We stress that, as can be noticed in Fig.~\ref{fig:other_filters}, to obtain the same result by using the P and the CP sequences, larger $n$s are required. 
For larger number of pulses the rate of change of $F_1^U(\omega,t_e,2n)$ for 
$\omega \lesssim 2 \pi/t_e$ becomes very small and the gate error is due 
to the fourth-order statistical properties of the noise in agreement with the results in Fig.~\ref{fig:figure:UDD-different-noise-cutoff}.

\begin{figure}[h!]
\centering
\includegraphics[width=0.6\textwidth]{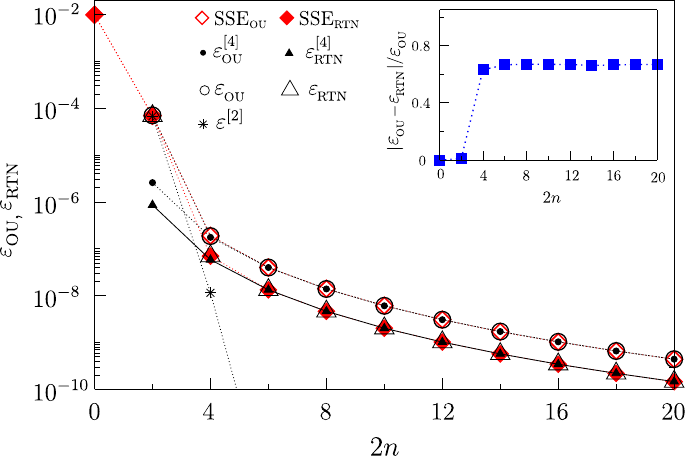}
\caption{Gate error under Uhrig DD at $t_e$ as a function  $n$, for $\omega_c = 5 \cdot 10^9$ rad/s. Symbols are the solution of the SSE: for  Gaussian noise (OU, open diamonds) and non-Gaussian noise (RTN, filled diamonds). Both processes have zero average
and the same variance, $\Sigma_\zeta=10^9$ rad/s. The Gaussian noise
is obtained by an ensemble of 256 RTNs with the same $\gamma= 1$ s$^{-1}$, whereas non-Gaussian noise is produced by a single RTN  with $\gamma= 1$ s$^{-1}$. The error given
by the second-order term in Eq.~\eqref{eq:error_formula} (stars), the contribution to the errors given by the Gaussian fourth-order term 
(second row in Eq.~\eqref{eq:error_formula}, filled circles for OU) and Gaussian plus non-Gaussian fourth-order terms (second and third row in Eq.~\eqref{eq:error_formula}, filled triangles for RTN),
and the total error given by second and fourth-order terms (open circles for OU, open triangles for RTN) are also shown. Inset: the difference between gate errors due to OU and RTN processes, normalized by the gate error due to OU noise; this highlights non-Gaussian effects evidenced by Uhrig DD.
}
\label{fig:Effetti_nonGauss}
\end{figure}

{\bf Non-Gaussian noise}  Uhrig's dynamical control capability to suppresses more efficiently the effect of second-order noise correlations 
makes this filter valuable for providing information of non-Gaussianity. To this end, we compare the effect of an RTN and an 
Ornstein–Uhlenbeck process (OU)~\cite{kb:papoulis} which have the same second-order statistics (and zero average value).
The error resulting from the exact numerical solution of the SSE is reported in  Fig.~\ref{fig:Effetti_nonGauss} and compared with the 
analytical approximation  Eq.~\eqref{eq:error_formula}.
Open diamonds correspond to OU and filled diamonds to RTN. 
The error due to second-order statistics almost vanishes for $n\ge 4$, and the curves obtained by SSE are captured by the fourth-order contributions in Eq.~\eqref{eq:error_formula}.

The error due to the OU process is entirely captured by the Gaussian fourth-order term in Eq.~\eqref{eq:error_formula}, while in the error due to the RTN both Gaussian and non-Gaussian terms contribute.
To highlight fourth-order statistic non-Gaussian effects in $\varepsilon$, in the inset of Figure~\ref{fig:Effetti_nonGauss} 
we plot the difference between the errors due to OU (open diamonds) and RTN (filled diamonds).  We observe that non-Gaussian fourth-order effects are evident already for $2n\ge 4$.

{\bf Spatially-correlated processes}
\label{sec:spatial-corr}
Dynamical control can also be used as a sensitive probe of noise correlations between processes affecting the two qubits an issue whose importance emerged in recent experiments~\cite{Yoneda2023}. 
Here we consider spatially-correlated processes\cite{ka:208-darrigo-njp-crosscorrelations}, and assume that  $z_\alpha(t)$ have the same statistical properties.  
Under these conditions (see Suppl. \ref{supp:stochastic}), 
spatial correlations are quantified by a single correlation coefficient~\cite{kb:papoulis}
\begin{equation}
    \mu=\frac{S_{z_1 z_2}(\omega)}{\sqrt{S_{z_1}(\omega)S_{z_2}(\omega)}},
    \label{eq:correlation_coefficient}
\end{equation}
that can be detected by spectral analysis.
It can be demonstrated along the same lines leading to Eq.~\eqref{eq:error_formula} that the gate error reads
\begin{equation}
    \varepsilon(\mu) = 2(1-\mu)\Big[\varepsilon^{[2]} + (1-\mu)\big(2 \varepsilon^{[4]}_g + \varepsilon_{ng}^{[4]}\big)\Big] \equiv \varepsilon^{[2]} (\mu) + \varepsilon_g^{[4]} (\mu)+\varepsilon_{ng}^{[4]}(\mu).
    \label{error_formula-correlations}
\end{equation}
where the $\varepsilon^{[2]}$, $\varepsilon_{g}^{[4]}$ and $\varepsilon^{[4]}_{ng}$ are given in Eq.~\eqref{eq:error_formula}.

\begin{figure}[h!]
\centering
\includegraphics[width=0.6\textwidth]{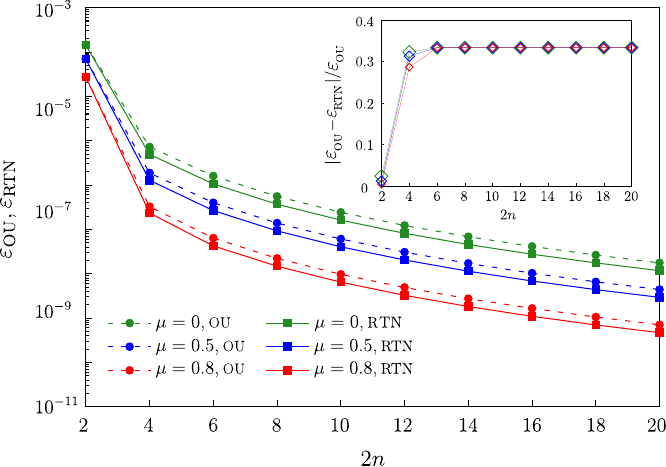}
\caption{Gate error under U in the presence of Gaussian (OU) and non-Gaussian (RTN) correlated noise versus the number of pulses. The symbols are the analytical result \eqref{error_formula-correlations} (circles for OU, squares for RTN), and lines are guides for the eye. 
Different colours represent different correlation coefficients $\mu=0, 0.5,0.8$ (green, blue, red). 
Inset: Difference between Gaussian and non-Gaussian gate errors scaled with the Gaussian error for different correlation coefficients.
 }
\label{fig:correlations}
\end{figure}

In Fig.~\ref{fig:correlations} we show $\varepsilon(\mu)$. 
The symbols (squares for OU, dots for RTN) are the analytic form Eq.~\eqref{error_formula-correlations} (reproducing the numerical solution of the SSE, not shown). 
For two pulses the error is due to second-order statistics $\varepsilon(\mu)\,\approx\,\varepsilon^{[2]}(\mu)$, therefore it does not distinguish the OU process from RTN. The difference between the errors is entirely due to noise correlations entering the pre-factor $2(1-\mu)$. 
For larger number of pulses, the error is dominated by fourth-order statistics and non-Gaussian
effects appear (difference between the squares and dots pairs for each colour). It is seen that the impact of correlation depends on Gaussianity as emphasized in the inset of Fig.~\ref{fig:Effetti_nonGauss} where we plot $ [\varepsilon_\text{OU}(\mu) -  \varepsilon_\text{RTN}(\mu)]/ \varepsilon_\text{OU}(\mu)$. For $2n=4$ the error is due to both second and fourth-order correlators which have a different dependence on $\mu$, for a larger number of pulses the errors are given by fourth-order correlators, resulting in a scaled difference between errors 
independent on $\mu$. 

\section{Discussion}
\label{sec:discussions}
In this work, we have studied the protection of coherence by DD during processing in a non-trivial quantum gate. This is an important issue for two-qubit gates whose duration is typically much longer than single-qubit ones. To this end, we tackled the problem of selective cancelling of non-commuting entries of the Hamiltonian, extending to transverse noise methods introduced for analyzing pure-dephasing longitudinal noise.

In particular, for the Ising-$xx$ coupling Hamiltonian, we studied pulsed control in the presence of {\em local} longitudinal noise focusing on the $W$-subspace where noise is transverse to the projected Hamiltonian. 
From a complementary perspective, the two-qubit "principal system" may probe characteristics of environmental noise.  We have shown that a QS protocol based on DD during entanglement generation may provide non-trivial information on the noise statistics, as on spatial noise correlations and/or on the fourth-order cumulant of the resulting stochastic process. Our result 
leverages generalized FFs describing DD while processing which differ from the standard FFs for longitudinal noise. Generalized FFs filter almost uniformly up to frequencies $~\sim \omega_c$ already with a small number of pulses yielding a very efficient protocol for QS of environmental noise. 

In particular, we suggest a simple procedure to extract relevant information on low-frequency longitudinal noise on each qubit of an entangling gate in a fixed coupling scheme. The noise variances $\Sigma_z^2$ can be extracted from each qubit coherence, which in the presence of quasi-static pure dephasing noise decays with a peculiar Gaussian law
$\rho_{+-}(t) \propto \exp{(- \Sigma_z^2 t^2)}$~\cite{falci2005PRL,ka:212-chiarellopaladino-njp-decoherence}, as it is observed in Ramsey experiments.
Then from an entangling gate operation, the presence of spatial correlations of the noise can be checked from the gate error under the P sequence.  In fact, for quasi-static noise, the analysis of Section 2.2 can be extended to correlated noise leading to $\epsilon_{qs}^{(P)} \simeq  \frac{\pi^2}{2^6} 2(1-\mu)   \left( \frac{\Sigma_\zeta}{\omega_c} \right)^2  n^{-2}$.
Due to higher order filtering properties, Uhrig dynamical control on the two-qubit gate may be employed for distinguishing quasi-static Gaussian noise, leading to an error scaling as $\epsilon_{qs}^{(U)} \propto 4(1-\mu)^2  \left( \frac{\Sigma_\zeta}{\omega_c} \right)^4 n^{-3.6}$, from quasi-static non-Gaussian noise. In fact, the results in Fig. \ref{fig:Effetti_nonGauss} for a RTN and Eq. \eqref{error_formula-correlations} indicate that the error scales as $\epsilon \propto 2(1-\mu)^2  \left( \frac{\Sigma_\zeta}{\omega_c} \right)^4 n^{-3.6}$. The quantitative distinction between the two processes requires the evaluation of the prefactors which may depend on the specific non-Gaussian process in the considered experimental setup. 

Since our results apply directly to single-qubit devices sensitive to low-frequency transverse noise, as the first generation of solid-state qubits~\cite{Vion2002}, properly designed single-qubit devices could work as quantum sensors of trispectrum if biased to make the leading noise transverse. The mapping into a single-qubit problem also suggests that there may be cases where anti-Zeno behaviour~\cite{facchi_unification_2004}
could manifest. Therefore a more complex scenario would emerge in the two-qubit dynamics where DD may be detrimental to the accuracy of entangling gates, possibly requiring quantum control and machine learning methods~\cite{ka:22-giannelli-pla-tutorial,ka:21-brown-njp-reinforcement}. 

We finally comment on the gate model we have chosen and on the relevance of different noise contributions. The $xx$-Ising interaction between qubits is a physical description of several implementations of quantum gates with fixed coupling, as capacitively or inductive coupled superconducting qubits~\cite{Kjaergaard_2020} or semiconducting qubits~\cite{chatterje2021},
as well as an effective description of cavity-mediated interactions. In these cases, local longitudinal noise is potentially the major semiclassical source of dephasing~\cite{Paladino2010}  which justifies our choice Eq.~\eqref{eq:H}, local transverse semiclassical noise, described by the Hamiltonian $\delta \mathcal{H}(t) = -\frac{1}{2} x_{1}(t) \, \sigma_{1x}  \otimes \mathbbm{1}_{2} - \frac{1}{2} \mathbbm{1}_{1} \otimes x_{2}(t) \, \sigma_{2x}\,$ being less relevant for our work. Indeed, low-frequency 
components would produce weak "transverse" dephasing between the invariant subspaces. The main effect would be leakage from the $W$-subspace, which should properly be described by a quantum noise model \cite{D'Arrigo_2012,NesiPRB2007,NesiNJP2007} outside this work's scope. In any case, the sequences of $\sigma_{\alpha y}$ pulses we consider tend to cancel also the $\sigma_{ \alpha x}$ coupling with the environment making the associated semiclassical noise less relevant, as we checked with SSE. Finally, noise directly affecting the $xx$ qubit coupling term is not expected to be important for fixed-coupling design or for qubits coupled via a transmission line since 
it would be longitudinal in the $W$ subspace. On the contrary, it may be non-negligible when the qubit coupling is implemented by a switchable circuit~\cite{McCourt2023}.

\section{Methods}\label{sec:methods}
\subsection{Open loop control}
\label{sec:openloop}
We consider a dynamical control operated by instantaneous pulses 
acting locally and simultaneously on each qubit. 
The control sequence aims to reduce the effect of fluctuations while performing an entangling gate operation.  
These two requirements may be fulfilled by a sequence of an even number of simultaneous $\pi$-pulses around the $y$-axis of the Bloch sphere of each qubit which tends to average out individually all the terms of ${\cal H}(t)$ Eq.~\eqref{eq:H} but the qubit-coupling.
It is described by the control 
\begin{equation}
\label{eq:control-sequence}
	{\cal H}_C(t)\,=\, {\cal V}_1(t)\otimes\mathbbm{1}_{2}\, + \,\mathbbm{1}_{1}\otimes{\cal V}_2(t), \quad \text{with} \quad {\cal V}_\alpha(t)= 
	-i \sum_{i=1}^{2n} \delta(t-t_i) \sigma_{\alpha y} 
\end{equation}
where the $\delta$-function approximates the process when the duration of the individual pulse is much smaller than the evolution time of the system under ${\cal H}$. 
Notice that the control Eq.~\eqref{eq:control-sequence} tends to suppress dynamically 
also local transverse noise coupled to each qubit via $\sigma_{\alpha x}$, thus we consider 
\begin{equation}
\delta \mathcal{H}(t) = 
-\frac{1}{2} \Big(x_{1}(t) \, \sigma_{1x} \,+\, z_{1}(t) \, \sigma_{1z}\Big) \otimes \mathbbm{1}_{2} - \frac{1}{2} \mathbbm{1}_{1} \otimes  \Big(x_{2}(t) \, \sigma_{2x}\,+\, z_{2}(t) \, \sigma_{2z}\Big)  \,.
\label{eq:noise}
\end{equation}
Under these conditions, we express the density matrix $\rho(t)$ of the system as a path-integral
over the realizations of the stochastic process. 
Denoting by $\rho\big(t | \vec \xi(t)\big)$ the density matrix associated with a single realization of the stochastic process $\vec \xi(t)=\{x_1(t),z_1(t), x_2(t), z_2(t)\}$ we write 
\begin{equation}
\rho(t) = \int {\mathcal D}[\vec \xi(t)] \,  P[\vec \xi(t)] \;  \rho\big(t | \vec \xi(t)\big) \, ,
	\label{eq:rho_t}
\end{equation} 
where $P[\vec \xi(t)]$ is the probability density for the realization $\vec \xi(t)$ of the noise.

For a given realization $\vec \xi(t)$, the dynamics generated by a sequence of two pulses alternated by two Hamiltonian evolutions for a time $\Delta t_i = t_{i+1}-t_i$ is described by the propagator
\begin{equation}
	{\cal U}(t_{i+1},t_{i-1}|\vec{\xi}(t))={\cal S}\,\hat{T}e^{-i\int_{t_i}^{t_{i+1}}{\cal H}(t^\prime)dt^\prime}\,{\cal S}\,
    \hat{T} e^{-i\int_{t_{i-1}}^{t_{i}}{\cal H}(t^\prime)dt^\prime} \, .
\label{eq:-z1}
\end{equation}
where ${\cal S}\equiv -\sigma_{1y}\otimes \sigma_{2y}$ and $\mathcal{H}(t) = \mathcal{H}_0 + \delta \mathcal{H}(t) $. 

For equally spaced pulses ($\Delta t_i \equiv \Delta t \ \ \forall i$)
the gate operation is not altered to leading order in $\Delta t$, 
provided that $\Delta t \ll \min\{{\tau}_{\xi\alpha}\}$, being
$\tau_{\xi_\alpha}$ the dominant (shortest)
correlation time associated with the noise $\xi_\alpha(t)$. 
Under these conditions, the noise can be approximated, for any $t \in [t_{i-1}, t_{i+1}]$, as a static stray component\cite{falci2005PRL} $\xi(t) \approx \xi$. Consequently, the integral simply factorizes $\int_i^{i+1} {\cal H}(t^\prime) dt^\prime \sim {\cal H}(t_i) \Delta t$. By expanding the exponential in Eq.~\eqref{eq:-z1} the evolution operator at the first order in $\Delta t$ reads
\begin{equation}
\begin{aligned}
	{\cal U}(t_{i+1},t_{i-1}|\vec{\xi}(t))
 &\simeq \mathbbm{1} + i \, {\omega_c} \, \Delta t \, \sigma_{1x}\otimes \sigma_{2x}  
   \simeq e^{i\, \omega_c \Delta t\, \sigma_{1x}\otimes \sigma_{2x} }
   \end{aligned}
\label{eq:-a2}
\end{equation}
Therefore the first order in $\Delta t$  ${\cal U}(t)$ implements a $\sqrt{i -\text{SWAP}}$ at time $t_e = 2 \Delta t$, noise effects being averaged out by the sequences of two pulses ${\cal S}$.
This result extends to any sequence of an even number $2n$ of pulses such 
that $\sum_{k=1}^{2n} \Delta t_k= t_e \ll \min\{{\tau}_{\xi_\alpha}\}$.

We used the error $\varepsilon$ Eq.~\eqref{eq:error_formula} in the 
entanglement generation protocol $\ket{+-} \to \ket{\psi_e}$ as a tool for noise sensing. Moreover, 
for $|z_i|<\omega_c$ the error $\varepsilon$ is close 
to the $W$-space infidelity thus it also quantifies the 
performance of DD in noisy gate processing.  

Notice that while the Hamiltonian ${\cal H}_0$ operates in the proper limit a $\sqrt{i -\text{SWAP}}$ gate (see supplemental \ref{supp:ham}), 
the gate under the $\cal S$ pulse sequence tends to preserve the ideal dynamics in the $W$-subspace and not in the $Z$-subspace. This is not a problem since 
the unitary Eq.~\eqref{eq:-a2} when acting on the $Z$ subspace can generate maximally entangled states. Therefore the DD sequences we consider preserve the ability of processing a perfectly entangling gate while decoupling. From the point of view 
of QS, the effective dynamics in the $Z$-subspace under the pulse sequences provide asymptotically information on the stochastic process $z_1 + z_2$ coupled transversally to the effective Hamiltonian.

\subsection{Dynamical control of pure dephasing correlated noise}
\label{sec:dd-puredephasing}
Here we focus on local longitudinal noise, and suppose that $z_1(t)$ and $z_2(t)$ are distinct stochastic processes with a correlation degree quantified by $\mu$, Eq.~\eqref{eq:correlation_coefficient}.
Control pulses $\mathcal{S}$ transform to $\pi$-rotation along the $z$-axis with propagator  ${\cal S} = \tau_{z}$, such that $\tau_z\, \hat{T}e^{-\frac{i}{2}\int_{t_{k}}^{t_{k+1}}  \mathcal{H}_g(t^\prime) dt^\prime}\,\tau_z=
\hat{T}e^{-\frac{i}{2}\int_{t_{k}}^{t_{k+1}} [-{\zeta}(t^\prime)\tau_x-\omega_c\tau_z]dt^\prime}$. 
Therefore, the effect of a control sequence can be included in the controlled-gate Hamiltonian
\begin{eqnarray}
&& {\cal H}_{cg}(t)\,=\,- \frac{\bar{\zeta}(t)}{2}\tau_x\,-\, \frac{\omega_c}{2} \tau_z  =  {\cal H}_n(t) \, +\, {\cal H}_c,
\qquad \bar{\zeta}(t)=(-1)^{k+1} {\zeta}(t), \,\,\,t \in [t_{k-1}, t_{k}[
  \label{eq:controlled-gate-noisyH}
\end{eqnarray}
where 
$ {\cal H}_c=-\, \frac{\omega_c}{2}\, \tau_z$
and
${\cal H}_n(t)= - \frac{1}{2} \bar \zeta(t) \,\tau_x $.
Thus, for preparation in the single-excitation subspace, 
the coupled qubit evolution under local longitudinal noise and local DD 
is mapped to a driven pseudo-two-state system subject to transverse noise.

Introduced the propagator ${\cal U}_c(t) = e^{\frac{i}{2}\omega_c t \tau_z}$, we can write the Hamiltonian
\begin{equation}
	\tilde{\cal H}_n(t)\, =\, {\cal U}_c^\dag(t) {\cal H}_n(t) {\cal U}_c(t) = -\zeta(t)\big(\tau_x \cos\omega_ct\,+\,\tau_y\sin\omega_ct\big),
	\label{eq:n0-02}
\end{equation}
that generate the dynamics in the ``toggling'' frame. This dynamics is described by $\tilde{\mathcal U} (t_e| \zeta(t))\,=\hat{T}e^{i\int_0^{t_e}\tilde{\cal{H}}_n(t^\prime)dt^\prime} \, $.

The overall propagator can be written as
${\cal U}(t_e)={\cal U}_c(t_e)\tilde{\mathcal U} (t_e| \zeta(t))$ and we have $\rho(t_e) = \mathcal{U}(t_e) \rho(0) \mathcal{U}^\dagger(t_e)$. Exploiting the fact that $\rho(0) = \ket{+-}\bra{+-}$ we can write the gate error as
\begin{equation}
	\begin{aligned}
		\varepsilon= \,&1 - \,\bra{\psi_e} {\cal U}(t_e)\rho(0) {\cal U}^\dagger(t_e) \ket{\psi_e}\\
		= \,&1 - \,\bra{\psi_e} {\cal U}_c(t_e)\mathcal{\tilde{U}}(t_e|\zeta(t))\rho(0) \mathcal{\tilde{U}}^\dagger(t_e|\zeta(t)){\cal U}_c^\dagger(t_e) \ket{\psi_e}\\
		=\,&1 - \,|\bra{+-}\tilde{\mathcal U}(t_e|\zeta(t)) \ket{+-}|^2
	\end{aligned}
	\label{eq:fidelity-togglingframe}
\end{equation}
To find analytic expressions for the gate error, we proceed analogously to~\cite{Green2013} and express the time propagator in the toggling frame  by its Magnus expansion:
\begin{equation}
	\tilde{\mathcal U} (t_e| \zeta(t))\,=\,e^{\Omega_1(t_e)+\Omega_2(t_e)+ \dots},
	\label{eq:MagnusExpansion}
\end{equation}
where, for simplicity, we omit the dependence of $\Omega_\alpha(t)$ on $\zeta (t)$. The first two terms of the expansion Eq.\eqref{eq:MagnusExpansion} read
\begin{eqnarray}
	\Omega_1(t_e)\,&=&\,i\big [a_{1x}(t_e)\tau_x\,+\,a_{1y}(t_e)\tau_y\big], \label{a-1}\\
	\Omega_2(t_e)\,&=&\,i\,a_{2z}(t_e)\tau_z, \label{a-2}
\end{eqnarray}
where
\begin{equation}
	\begin{aligned}
		&a_{1x}(t_e) = \frac{1}{2}\int_0^{t_e} dt_1\,\zeta(t_1)\cos(\omega_ct_1) \,, \quad a_{1y}(t_e) = \int_0^{t_e} dt_1\,\zeta(t_1) \sin(\omega_ct_1) \, ,\\
		&a_{2z}(t_e) = \frac{1}{4}\int_0^{t_e} dt_1 \int_0^{t_1} dt_2 \zeta(t_1)\zeta(t_2)\sin(\omega_c(t_1-t_2)),\\
	\end{aligned}
\end{equation}
As discussed in App.~\ref{supp:magnus}, by truncating the Magnus expansion to the third-order 
the gate error can be approximated as follows
\begin{equation}
	\varepsilon \simeq \varepsilon^{[2]} + \varepsilon^{[4]},
\label{eq:error-formula}
\end{equation}
where $\varepsilon^{[2]} = \braket{a_{1y}^2}$ and $\varepsilon^{[4]} = \braket{a_{2z}^2}$ are of second and fourth order in the noise, respectively.
The validity of these approximations is confirmed by the results presented in Sec.~\ref{sec:results}.

\subsection{Filter function formalism}
\label{sec:filterfunction}
The gate error in Eq.~\eqref{eq:error-formula} can be expressed in terms of FFs of subsequent noise cumulants (for details c.f. supp.~\ref{supp:filter}), defined from the Uhrig filter~\cite{uhrig2007PRL} 
 \begin{equation}
y_n(\alpha,t_e) \equiv 1+(-1)^{n+1}e^{i\alpha t_e}\,+\,2\sum_{k=1}^{n}(-1)^k
e^{i\alpha t_k} \, 
\label{Uhrig_filter}
\end{equation}

As an example, the second-order contribution reads
\begin{equation}
\begin{aligned}
	\varepsilon^{[2]}&=\,
	\braket{\Big(\frac{i}{2}\int_0^{t_e} dt_1\,\bar \zeta(t_1) \sin(\omega_ct_1)\big]\Big)^2}\\
	&=\frac{1}{16}\int_{-\infty}^{+\infty}\frac{d\omega}{2\pi}S_{\zeta}(\omega)
\Big[\frac{\big|y_n(\omega+\omega_c,t_e)\big|^2}{(\omega+\omega_c)^2}
\,+\,\frac{\big|y_n(\omega-\omega_c,t_e)\big|^2}{(\omega-\omega_c)^2}\Big]\\
	&=\,\int_{-\infty}^{+\infty}\frac{d\omega}{2\pi}S_{\zeta}(\omega) F_{1}^{}(\omega,\omega_c,t_e,2n).
\label{eq:a1y}
\end{aligned}
\end{equation}
This expression defines the FF of second order $F_{1}^{}(\omega,\omega_c,t_e,2n)=\frac{\big|y_n(\omega+\omega_c,t_e)\big|^2}{(\omega+\omega_c)^2} +\,\frac{\big|y_n(\omega-\omega_c,t_e)\big|^2}{(\omega-\omega_c)^2}$. 

The same calculation for the forth order contribution reveals that $\varepsilon^{[4]}$ can be decomposed in a Gaussian $\varepsilon_g^{[4]}$ and a non-Gaussian $\varepsilon^{[4]}_{ng}$ contributions. Analogously to what is done for $\varepsilon^{[2]}$, we can define two additional FFs (whose explicit expression is left to supp.~\ref{supp:filter}) and write
\begin{equation}
	\varepsilon^{[4]}_g\,=\,\,\int_{-\infty}^{+\infty}\frac{d\omega_1}{2\pi}\,S(\omega_1) 
\int_{-\infty}^{+\infty}\frac{d\omega_2}{2\pi}\,S(\omega_2)\,
F_{2,g}(\vec{\omega_2},\omega_c,t_e,2n),
\label{eq:a2zg}
\end{equation}
and
\begin{eqnarray}
	\varepsilon^{[4]}_{ng}\,=\,
\,\int_{-\infty}^{+\infty}\frac{d\omega_1}{2\pi}\,
\int_{-\infty}^{+\infty}\frac{d\omega_2}{2\pi}\,
\int_{-\infty}^{+\infty}\frac{d\omega_3}{2\pi}\, 
S_{\zeta 3}(\vec{\omega_3}) F_{2,ng}(\vec{\omega_3},\omega_c,t_e,2n). 
\label{eq:a2zng}
\end{eqnarray}
and write finally Eq.~\eqref{eq:error_formula} for the gate error.

We remark that the above expression of the gate error holds for any decoupling sequence consisting of an even number of simultaneous $\pi_y$ pulses applied at times $t_k = \delta_k t_e$ with $0 < \delta_k < 1$ and $k \in {1, 2n}$, such that $t_{2n}=t_e$. 
Each sequence corresponds to different filters thus allowing either to (partly) cancel environmental effects
to various orders\cite{PazSilva2017} or to filter out relevant spectral components (filtering order).   

\begin{figure}[h!]
\centering
	\includegraphics[width=0.8\textwidth]{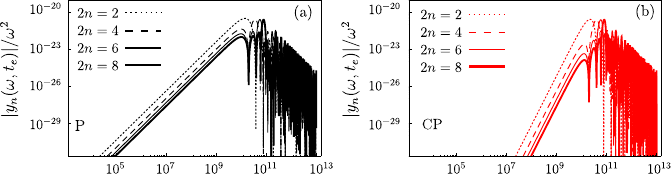}
\caption{Plots of the second-order filters $F_1(\omega,t_e, 2n)$ for P (left panel) and CP (right panel) as a function of $\omega$ and for $\omega_c=5\cdot 10^9$ rad/s.
In each figure, the different line styles correspond to different numbers of applied pulses. }
\label{fig:other_filters}
\end{figure}

In this work, we focus on three specific decoupling sequences, the Periodic (P), the Carr-Purcell (CP), and the Uhrig (U). They are characterized by different $\delta_k$ and, consequently, different pulse intervals: For the periodic sequence we have $\delta_k = k/2n$ and $\Delta t = t_e/2n$; for the Carr-Purcell we have $\delta_k = (k -1/2)/2n$ and $\Delta t_k = 2 \Delta t_1$ with $\Delta t_1 = \Delta t_{2n+1} = t_e/4n$; for the Uhrig case we have $\delta_k = \sin^2[\pi k/(2n+2)]$. Notice that in the limit of a two-pulses cycle ($n = 2$) the Uhrig sequence reduces to the Carr-Purcell one.

In Fig. \ref{fig:other_filters} we show the second order Urhig filters $F_1(\omega,t_e, 2n) = \frac{\big|y_n(\omega, t_e)\big|^2}{\omega^2}$
for the P and CP sequences, $F_1(\omega,t_e, 2n)$ for the Uhrig sequence is reported in Fig. \ref{fig:Urhig_filters}.

{\bf Acknowledgements}
This work was supported by the  PNRR MUR project PE0000023-NQSTI, by the 
QuantERA grant SiUCs (Grant No. 731473), by the University of Catania, Piano Incentivi Ricerca di Ateneo 2020-22, project Q-ICT. EP
acknowledges the COST Action CA 21144 superqumap.

{\bf Data availability statement}
The data that support the ﬁndings of this study are available from the corresponding author upon request.

{\bf Author Contributions}
A.D. and E.P. conceived the model and carried out the analytical calculations; A.D. carried out the numerical calculations; A.D., G.F., G.P. and E.P. analyzed the data. All authors discussed the physics and contributed to writing the manuscript.

{\bf Competing interests} The authors declare no competing interests.

\newpage
\appendix

\section{Two-qubit entangling gate}\label{supp:ham}
We consider two qubits labelled by $\alpha = 1,2$
living in the tensor product of the two-dimensional Hilbert space $H_1 \otimes H_2$. We define $\sigma_{\alpha k}$, with $k = x, y, z$, the Pauli matrices acting on the qubit $\alpha $. The factorized computational basis $\{\ket{\mu \nu} := \ket{\mu}_1 \otimes \ket{\nu}_2,\; \mu, \nu = \pm \}$ is such that $\sigma_{\alpha  z}\ket{\pm}_\alpha = \mp \ket{\pm}_\alpha$. The qubits are coupled by an Ising $x-x$ interaction. We considered the Hamiltonian 
$$
{\cal H} = -\frac{\Omega_1 + z_1(t)}{2} \, \sigma_{1z} \otimes \mathbbm{1}_{2}
 -\frac{\Omega_2 + z_2(t)}{2}  \, \mathbbm{1}_{1} \otimes \sigma_{2z}
+ \frac{\omega_c}{2} \, \sigma_{1x} \otimes \sigma_{2x}  
-\frac{x_{1}(t)}{2} \, \sigma_{1x} - \frac{x_{2}(t)}{2} \sigma_{2x}.
$$
where terms describing both longitudinal and transverse noise affecting each qubit appear. 
In the absence of transverse noise ($x_\alpha =0$) 
the Hamiltonian $\cal H$ is block diagonal the Hilbert space being the direct sum of two invariant subspaces denoted by $W= \mathrm{span}\{\ket{+-}, \ket{-+}\}$ and 
$Z= \mathrm{span}\{ \ket{++}, \ket{--}\}$. 
If the qubits are identical or if $|\Omega_1-\Omega_2| \ll \omega_c$ we obtain  an effective ${\cal H}_0$ with the structure 
of Eq.~\eqref{eq:H} where 
$\Omega= \Omega_1 + \Omega_2$ and $\omega_c \to \omega_c + |\Omega_1-\Omega_2|^2/(2 \omega_c)$ whose eigenvalues and eigenvectors 
are reported in Tab.~\ref{tab:eigensystem0}. This Hamiltonian implements the entanglement-generation operation Eq.~\eqref{eq:entanglement-generation} studied in this work.  

In the usual limit $\omega_c \ll \Omega$ the Hamiltonian in the $Z$ subspaces presents only renormalized diagonal entries thus in this limit ${\cal H}_0$ can implement a gate locally equivalent to $\sqrt{{\rm i-SWAP}}$ by evolving the system for a time $t_e = \pi/2\omega_c$. 

\begin{table}[t!]
\centering
{\renewcommand{\arraystretch}{1.4}
\begin{tabular}{|l|ll|}
\hline
$\beta$ & $\omega_\beta$ & $\ket{\beta}$ \\
\hline
0 & $- \sqrt{\Omega^2 + (\omega_c/2)^2}$ & 
$-  (\sin \vartheta/2  )\vert ++ \rangle+ (\cos \vartheta/2 ) \vert -- \rangle$ \\
1 & $-\omega_c/2$ & $\big[\vert +- \rangle - \vert -+ \rangle \big]/\sqrt{2}$ \\
2 & $\phantom{-}\omega_c/2$ & $\big[\vert +- \rangle + \vert -+ \rangle \big]/\sqrt{2}$ \\
3 & $\phantom{-}\sqrt{\Omega^2 + (\omega_c/2)^2}$ & 
$\cos (\vartheta/2 )\vert ++ \rangle + \sin (\vartheta/2) 
\vert -- \rangle $ \\
\hline
\end{tabular}
}
\caption{
\label{tab:eigensystem0} 
Eigenvalues and eigenvectors of ${\mathcal{H}}_0 $. Here $\tan \vartheta = - \omega_c/(2 \Omega)$. The two Hilbert subspaces are spanned by  $\{ |1 \rangle, |2 \rangle\}$ and $\{ |0 \rangle, |3 \rangle\}$
}
\end{table}

\section{Characterization of the stochastic processes}
\label{supp:stochastic} 
We considered noise described by two stochastic processes
$\xi_\alpha \equiv \{ x_\alpha(t), z_\alpha(t) \}$ assumed to be 4-th order stationary and with vanishing  average, $\braket{\xi_\alpha(t)}=0$, 
where $\langle \cdot \rangle $ indicates the ensemble average.
The lowest order correlation functions are $C_{\xi_\alpha}(t+\tau, t)= \braket{\xi_\alpha(t+\tau)\xi_\alpha(t)}  \equiv  C_{\xi_\alpha}(\tau)$
and the cross-covariance $C_{\xi_1 \xi_2}(\tau) \, = \, \langle \xi_1(t+\tau)\xi_2(t)\rangle - \overline{\xi}_1\overline{\xi}_2\,$~\cite{kb:papoulis}.
The power spectrum $S_{\xi_\alpha}(\omega)$ and the cross-spectrum $S_{\xi_1 \xi_2}(\omega)$ are given by 
\begin{equation}
S_{\xi_\alpha}(\omega) = \int_{-\infty }^\infty d \tau \, 
	C_{\xi_\alpha}(\tau) \, e^{i \omega \tau}, \qquad
S_{\xi_1 \xi_2}(\omega) = \int_{-\infty }^\infty d \tau \, 
C_{\xi_1 \xi_2}(\tau) \, e^{i \omega \tau} \, .
\label{def_spectrum}
\end{equation}
In general, the amount of correlation between two stochastic processes $\eta_i(t)$ is quantified by the correlation factor 
\begin{eqnarray}
   \mu=\frac{\langle [\eta_1(t)-\overline{\eta}_1][\eta_2(t)-\overline{\eta}_2]\rangle}
       {\sqrt{\langle[\eta_1(t)-\overline{\eta}_1]^2\rangle
              \langle[\eta_2(t)-\overline{\eta}_2]^2\rangle}} \, ,
\label{XCorrelatioFactor}
\end{eqnarray}
where $\overline{\eta}_i \equiv \langle \eta_i(t)\rangle $.
Here we assume that each component of $\xi_\alpha$ is the sum of independent fluctuating factors having the same variance, for instance, $z_\alpha (t) = \sum_k c_k \delta z_{\alpha,k}(t)$ where the variance of $\delta z_{\alpha,k}(t)$ does not depend on $k$, $\Sigma_{z,\alpha}$.
This assumption is not restrictive, for instance, it models spatially-correlated processes and cross-talk effects in coupled transmons ~\cite{ka:208-darrigo-njp-crosscorrelations} or flux noise correlations between two loops of a tunable flux qubit~\cite{gustavsson2012PRL}
or tunable capacitively-shunted flux qubits~\cite {Trappen2023}, possibly due to non-local sources of flux noise or junction critical current noise.
Under these conditions, the degree of correlations is expressed by the correlation coefficient relating the cross-spectrum to the individual
power spectra and detectable  by spectral analysis
\begin{equation}
	S_{\xi_1 \xi_2}(\omega) = \mu \,\sqrt{S_{\xi_1}(\omega)S_{\xi_2}(\omega)} \, .
\label{Spectrum-correlation factor}
\end{equation}

To point out non-Gaussian effect we also evaluate the first non-vanishing higher-order correlator, i.e. the fourth-order cumulant $C_{\xi}^{(4)}(\vec \tau_3) = \braket{\braket{\xi(t_1) \xi(t_2) \xi(t_3) \xi(t_4)}}$, where $\vec \tau_3 = (\tau_1, \tau_2, \tau_3)$ with $\tau_i = t_{i+1}- t_1$, and the trispectrum 
\begin{equation}
    S_{\xi 3}(\vec \omega_3) = \int_{-\infty}^{\infty} d \vec \tau_3  e^{- i \vec\omega_3 \cdot \vec\tau_3} \, C^{(4)}_{\xi}(\vec \tau_3),
    \label{eq:trispectrum}
\end{equation}
where $\vec \omega_3 = (\omega_1, \omega_2, \omega_3)$.

\section{Magnus expansion}\label{supp:magnus}
In this appendix, we discuss some details of the derivation of the error in the Eq.~\eqref{eq:error-formula}. 
To this purpose, we notice that the third term of the Magnus expansion in Eq.~\eqref{eq:MagnusExpansion} reads
\begin{equation}
    \Omega_3 = i [a_{3x}(t_e) \tau_x + a_{3y}(t_e) \tau_y], 
\end{equation}
with
\begin{equation}
	\begin{aligned}
		&a_{3x}(t_e) = \frac{1}{12}\int_0^{t_e} dt_1\int_0^{t_1} dt_2\int_0^{t_2} dt_3\,        
    \zeta(t_1)\zeta(t_2)\zeta(t_3)  
\Big[\cos\omega_c(t_1-t_2+t_3)-\frac{1}{2}\cos\omega_c(t_1+t_2-t_3) -\cos\omega_c(t_1-t_2-t_3)\Big],  \\
&a_{3y}(t_e) =  \frac{1}{12}\int_0^{t_e} dt_1\int_0^{t_1} dt_2\int_0^{t_2} dt_3\,                        
    \zeta(t_1)\zeta(t_2)\zeta(t_3) \Big[-\sin\omega_c(t_1-t_2+t_3) +\frac{1}{2}\sin\omega_c(t_1+t_2-t_3) 
-\sin\omega_c(t_1-t_2-t_3)\Big] \, .
\end{aligned}
\end{equation} 
By truncating the Magnus expansion Eq.~\eqref{eq:MagnusExpansion} to the third order, the gate error at $t_e$ reads
\begin{equation}
\varepsilon \simeq \langle a_{1y}^2\rangle+\,\langle a_{2z}^2\rangle\,+\,
\langle 2a_{1y}a_{3y} \rangle\,-\,\frac{1}{3}\langle a_{1y}^4\rangle\,-\,
\frac{1}{3}\langle a_{1y}^2a_{1x}^2\rangle.
\label{eq:error-aterms}
\end{equation}
By comparing the maximum value of each contribution to the gate error, it is possible to verify that the last three terms are negligible
to the first two justifying the approximation  Eq.~\eqref{eq:error-formula} for the gate error.
This is also confirmed by the exact numerical solution of the SSE for the considered pulse sequences. 

\section{Derivation of the filter functions}\label{supp:filter}
In this appendix, we report the details of the derivation of the FFs. 
We first observe that the effect of dynamical control in the SWAP subspace is that of decomposing the time evolution in Eqs.~\eqref{a-1}-\eqref{a-2} as
\begin{equation}
\int_0^{t_e}dt_1\,{\bar \zeta}(t_1)\ldots=\sum_{k=1}^{n+1}\int_{t_{k-1}}^{t_k}dt_1 \,(-1)^k\, \zeta(t_1)\ldots.
\label{eq:barnoise}
\end{equation} 
Let us substitute this decomposition in Eq.~\eqref{eq:a1y}.
\begin{equation}
	\begin{aligned}
		\varepsilon^{[2]} &=\,\braket{\Big(
		\frac{i}{2}\int_0^{t_e} dt_1\,\bar \zeta(t_1) \sin(\omega_ct_1)\big]\Big)^2}\\
&=-\frac{1}{4}\int_0^{t_e} dt_1\int_0^{t_e}dt_2\,\sin\omega_ct_1\,\sin\omega_ct_2 \,
\big\langle \bar \zeta(t_1)\,\bar \zeta(t_2)\big\rangle\\
&=-\frac{1}{4}\sum_{k=1}^{n+1}\int_{t_{k-1}}^{t_k}dt_1 \,(-1)^k
\sum_{j=1}^{n+1}\int_{t_{j-1}}^{t_j}dt_2 \,(-1)^j\,\sin\omega_ct_1\sin\omega_ct_2 \,
\big\langle \zeta(t_1)\, \zeta(t_2)\big\rangle\\
&=-\frac{1}{4}\int_{-\infty}^{+\infty}\frac{d\omega}{2\pi}S_{\zeta}(\omega)
\sum_{k,j=1}^{n+1}(-1)^{k+j}\int_{t_{k-1}}^{t_k}dt_1 \,
\int_{t_{j-1}}^{t_j}dt_2 \,
\sin(\omega_ct_1)\sin(\omega_ct_2) \,e^{i\omega(t_1-t_2)}\\
&=-\frac{1}{4}\int_{-\infty}^{+\infty}\frac{d\omega}{2\pi}S_{\zeta}(\omega)
\sum_{k,j=1}^{n+1}(-1)^{k+j}\int_{t_{k-1}}^{t_k}dt_1 \,
\int_{t_{j-1}}^{t_j}dt_2 \Big[\cos\omega_c(t_1-t_2)\,-\,\cos\omega_c(t_1+t_2)\Big]\, 
\,e^{i\omega(t_1-t_2)},
	\end{aligned}
\label{eq:aind}
\end{equation}
where we used that by definition
\begin{equation}
\langle \zeta(t_1)\zeta(t_2)\rangle\,=\,\int_{-\infty}^{+\infty}
\frac{d\omega}{2\pi}\,e^{i\omega(t_1-t_2)}S_{\zeta}(\omega).
\label{eq:power_spectrum}
\end{equation}
We notice that 
\begin{equation}
\sum_{k=1}^{n+1}(-1)^{k}\int_{t_{k-1}}^{t_k}dt_1\,e^{i\alpha t_1}\,=\,
\frac{1}{i\alpha}\big[1+(-1)^{n+1}e^{i\alpha t_e}\,+\,2\sum_{k=1}^{n}(-1)^k
e^{i\alpha t_k}\big]=\frac{1}{i\alpha}y_n(\alpha,t_e),
\label{Uhrig_filterA}
\end{equation}
being $y_n(\alpha, t_e)$ the Uhrig filter introduced in Eq.~\eqref{Uhrig_filter},
Therefore we can compute the time integrals in the last row of Eq.~\eqref{eq:aind} due to $\cos\omega_c(t_1-t_2)$ as follows
\begin{equation}
	\begin{aligned}
&\sum_{k,j=1}^{n+1}(-1)^{k+j}\int_{t_{k-1}}^{t_k}dt_1 \,
\int_{t_{j-1}}^{t_j}dt_2 \,
\cos\omega_c(t_1- t_2)\,e^{i\omega(t_1-t_2)}\\
=\frac{1}{2}&\sum_{k,j=1}^{n+1}(-1)^{k+j}\Bigg[
\int_{t_{k-1}}^{t_k}dt_1 \,e^{i(\omega+ \omega_c)t_1}
\int_{t_{j-1}}^{t_j}dt_2 \,e^{-i(\omega+\omega_c)t_2}\,+\,
\int_{t_{k-1}}^{t_k}dt_1 \,e^{i(\omega- \omega_c)t_1}
\int_{t_{j-1}}^{t_j}dt_2 \,e^{-i(\omega-\omega_c)t_2}\Bigg]\\
=\frac{1}{2}&\Big[\frac{1}{(\omega+\omega_c)^2}\big|y_n(\omega+\omega_c,t_e)\big|^2
		\,+\,\frac{1}{(\omega-\omega_c)^2}\big|y_n(\omega-\omega_c,t_e)\big|^2\Big].
	\end{aligned}
\end{equation}
Analogous calculations for the contribution due to $\cos \omega_c(t_1+t_2)$ leads to
\begin{equation}
\sum_{k,j=1}^{n+1}(-1)^{k+j}\int_{t_{k-1}}^{t_k}dt_1 \,
\int_{t_{j-1}}^{t_j}dt_2 \,
\cos\omega_c(t_1+t_2)\,e^{i\omega(t_1-t_2)}
= \frac{1}{\omega^2-\omega_c^2} \,\Re[y_n(\omega+\omega_c,t_e)\,y_n^*(\omega-\omega_c,t_e)] \, 
\label{eq:aiy2-04A}
\end{equation}
which is vanishing for all the considered sequences.
Combining all these terms we obtain obtain Eq.~\eqref{eq:a1y}.

To evaluate the fourth-order contribution in Eq.~\eqref{eq:error-formula}, we observe that
\begin{equation}
\int_0^{t_e} dt_1\,\int_0^{t_1} dt_2\,=\,
\sum_{k=2}^{n+1}\int_{t_{k-1}}^{t_k}dt_1\sum_{m=1}^{k-1}\int_{t_{m-1}}^{t_m}dt_2\,+\,
\sum_{k=1}^{n+1}\int_{t_{k-1}}^{t_k}dt_1\int_{t_{k-1}}^{t_1}dt_2,
\label{eq:a2z-doubleintegral-stepsA}
\end{equation}
and we introduce the integral operator
\begin{equation}
\begin{aligned}
\mathcal{A} = \,\frac{1}{16}
&\Bigg(
\sum_{k=2}^{n+1}(-1)^k\int_{t_{k-1}}^{t_k}dt_1\sum_{m=1}^{k-1}(-1)^m\int_{t_{m-1}}^{t_m}dt_2\,+\,
\sum_{k=1}^{n+1}\int_{t_{k-1}}^{t_k}dt_1\int_{t_{k-1}}^{t_1}dt_2\Bigg)\times\\
&\Bigg(
\sum_{j=2}^{n+1}(-1)^j\int_{t_{j-1}}^{t_j}dt_3\sum_{l=1}^{j-1}(-1)^l\int_{t_{l-1}}^{t_l}dt_4\,+\,
\sum_{j=1}^{n+1}\int_{t_{j-1}}^{t_j}dt_3\int_{t_{j-1}}^{t_3}dt_4\Bigg).
\end{aligned}
\label{eq:a2z-02A}
\end{equation}
Therefore we can write
\begin{equation}
	\begin{aligned}
		\varepsilon^{[4]}\,=\,&\Big\langle\Big(
\frac{i}{4}\int_0^{t_e} dt_1\,\int_0^{t_1} dt_2\,\bar \zeta(t_1)\bar \zeta(t_2) \sin\omega_c(t_1-t_2)\big]\Big)^2\Big\rangle\nonumber\\
	=	&\frac{1}{16}\int_0^{t_e} dt_1\,\int_0^{t_1} dt_2\int_0^{t_e} dt_3\,\int_0^{t_3} dt_4 
\langle \bar \zeta(t_1)\bar \zeta(t_2) \bar \zeta(t_3)\bar \zeta(t_4) \rangle 
\sin\omega_c(t_1-t_2)\sin\omega_c(t_3-t_4)\\
		=\,& {\cal A}\Big\{\langle \zeta(t_1) \zeta(t_2) \zeta(t_3) \zeta(t_4) \rangle 
\sin\omega_c(t_1-t_2)\sin\omega_c(t_3-t_4)\Big\} \, ,
	\end{aligned}
\label{eq:a2z-01}
\end{equation}
The average $\braket{\zeta(t_1) \zeta(t_2) \zeta(t_3) \zeta(t_4)}$ can be decomposed in a Gaussian and a non-Gaussian contributions
\begin{equation}
\begin{aligned}
\langle \zeta(t_1) \zeta(t_2) \zeta(t_3) \zeta(t_4) \rangle\,&=\, \braket{\zeta(t_1) \zeta(t_2) \zeta(t_3) \zeta(t_4)}_g + \braket{\zeta(t_1) \zeta(t_2) \zeta(t_3) \zeta(t_4)}_{ng}, \\
\end{aligned}
\label{eq:fourth_average}
\end{equation}
defining the Gaussian and non-Gaussian contribution to the gate error
\begin{equation}
	\begin{aligned}
		&\varepsilon_g^{[4]}\,=\, {\cal A}\Big\{
\langle \zeta(t_1) \zeta(t_2) \zeta(t_3) \zeta(t_4) \rangle_g\,
\sin\omega_c(t_1-t_2)\sin\omega_c(t_3-t_4)\Big\}\\
&\varepsilon^{[4]}_{ng}\,=\, {\cal A}\Big\{
\langle \zeta(t_1) \zeta(t_2) \zeta(t_3) \zeta(t_4) \rangle_{ng}\,
\sin\omega_c(t_1-t_2)\sin\omega_c(t_3-t_4)\Big\}.
	\end{aligned}
\end{equation}
Before continuing we report the explicit expression of the two contributions of the average 
\begin{equation}
\begin{aligned}
	&\langle \zeta(t_1) \zeta(t_2) \zeta(t_3) \zeta(t_4) \rangle_g \phantom{_n}  =\,
\langle \zeta(t_1) \zeta(t_2)\rangle\langle \zeta(t_3) \zeta(t_4) \rangle\,+\,
\langle \zeta(t_1) \zeta(t_3)\rangle\langle \zeta(t_2) \zeta(t_4) \rangle\,+\, 
\langle \zeta(t_1) \zeta(t_3)\rangle\langle \zeta(t_2) \zeta(t_4) \rangle\,\,\\
	&\langle \zeta(t_1) \zeta(t_2) \zeta(t_3) \zeta(t_4) \rangle_{ng}=\,
\langle\langle \zeta(t_1) \zeta(t_2)\rangle\langle \zeta(t_3) \zeta(t_4) \rangle\rangle.
\end{aligned}
\end{equation}
The Gaussian contributions in Eq.~\eqref{eq:fourth_average} can be expressed in terms of the noise power spectrum Eq.~\eqref{eq:power_spectrum}
\begin{equation}
\begin{aligned}
	\langle \zeta(t_1) \zeta(t_2) &\zeta(t_3) \zeta(t_4) \rangle_g=\,
\langle \zeta(t_1) \zeta(t_2)\rangle\langle \zeta(t_3) \zeta(t_4) \rangle\,+\,
\langle \zeta(t_1) \zeta(t_3)\rangle\langle \zeta(t_2) \zeta(t_4) \rangle\,+\, 
\langle \zeta(t_1) \zeta(t_3)\rangle\langle \zeta(t_2) \zeta(t_4) \rangle\,\,\\
=& \int_{-\infty}^{+\infty}\frac{d\omega_1}{2\pi}\,S_\zeta(\omega_1)
\int_{-\infty}^{+\infty}\frac{d\omega_2}{2\pi}\,S_\zeta(\omega_2)\Bigg[
e^{i\omega_1(t_1-t_2)}e^{i\omega_2(t_3-t_4)}\,+ e^{i\omega_1(t_1-t_3)}e^{i\omega_2(t_2-t_4)}\,+\,e^{i\omega_1(t_1-t_4)}e^{i\omega_2(t_2-t_3)}\Bigg]
\end{aligned}
\label{eq:4th-average-gaussian-process}
\end{equation}
On the other side, the non-Gaussian term can be expressed in terms of the  trispectrum \eqref{eq:trispectrum}
\begin{eqnarray}
\langle \zeta(t_1) \zeta(t_2) \zeta(t_3) \zeta(t_4) \rangle_{ng}=\,
\int_{-\infty}^{+\infty}\frac{d\omega_1}{2\pi}\,e^{i\omega_1(t_2-t_1)}
\int_{-\infty}^{+\infty}\frac{d\omega_2}{2\pi}\e^{i\omega_2(t_3-t_1)}
\int_{-\infty}^{+\infty}\frac{d\omega_3}{2\pi}e^{i\omega_3(t_4-t_1)}\,S_{\zeta 3}({\omega}_1,{\omega}_2,{\omega}_3).
\label{eq:4th-average-nongaussian-processA}
\end{eqnarray}
Collecting these results we obtain
\begin{equation}
\begin{aligned}
	\varepsilon^{[4]}_g\,=\,&\int_{-\infty}^{+\infty}\frac{d\omega_1}{2\pi}\,S(\omega_1)
\int_{-\infty}^{+\infty}\frac{d\omega_2}{2\pi}\,S(\omega_2)\,
F_{2,g}(\omega_1,\omega_2,\omega_c,t_e,2n), \\
	\varepsilon^{[4]}_{ng}\,=\,&\int_{-\infty}^{+\infty}\frac{d\omega_1}{2\pi}\,
\int_{-\infty}^{+\infty}\frac{d\omega_2}{2\pi}\,
\int_{-\infty}^{+\infty}\frac{d\omega_3}{2\pi}\,S_{\zeta 3}(\omega_1,\omega_2,\omega_3)
F_{2,ng}(\omega_1,\omega_2,\omega_3,\omega_c,t_e,2n),
\end{aligned}
\end{equation}
which defines the fourth-order Gaussian and non-Gaussian FFs $F_{2,g}(\omega_1,\omega_2,\omega_c,t_e,2n)$ and $F_{2,ng}(\omega_1,\omega_2,\omega_3,\omega_c,t_e,2n)$. 
To calculate these filters $F_{2, (n)g}$ we first introduce the function
\begin{equation}
\begin{aligned}
\chi_n(\alpha,\beta,t_e) &\equiv \Bigg(\sum_{k=2}^{n+1}(-1)^k\int_{t_{k-1}}^{t_k}dt_1\sum_{m=1}^{k-1}(-1)^m\int_{t_{m-1}}^{t_m}dt_2\,+\,
\sum_{k=1}^{n+1}\int_{t_{k-1}}^{t_k}dt_1\int_{t_{k-1}}^{t_1}dt_2\Bigg)
 e^{i\alpha t_1}e^{i\beta t_2}\,\\
&= \, (-1)^{n+2}\frac{e^{i\alpha t_e}}{\alpha\beta}\,y_n(\beta,t_e)\,-\,
2\sum_{k=1}^n(-1)^k\frac{e^{i\alpha t_k}}{\alpha\beta}\,\tilde{y}_{n,k}(\beta,t_e)\,+\,
\frac{e^{i(\alpha+\beta) t_e}-1}{\alpha(\alpha+\beta)},
\end{aligned}
\label{eq:a2z-04A}
\end{equation}
where $\tilde{y}_{n,k}(\beta,t_e)\,=\,1+(-1)^{k+1}e^{i\beta t_k}\,+\,2\sum_{m=1}^{k}(-1)^m e^{i\beta t_m}$,
being the $t_k$s the same entering in ${y}_{n}(\beta,t_e)$.
Therefore we have ${\cal A}\Big\{e^{i\alpha t_1}e^{i\beta t_2}e^{i\gamma t_3}e^{i\delta t_4}\Big\}\,=\,
\chi_n(\alpha,\beta,t_e)\chi_n(\gamma,\delta,t_e)$. 
Exploiting this result, we obtain the explicit expressions for the FFs
\begin{equation}
\begin{aligned}
F_{2,g}(\vec \omega_2,\omega_c)\,=\,\frac{1}{2^6}\big\{
|&\chi_n(\omega_1+\omega_c,\omega_2-\omega_c,t_e)|^2\,+\,
|\chi_n(\omega_1-\omega_c,\omega_2+\omega_c,t_e)|^2\\
+\,2\Re\big[&\chi_n(\omega_1+\omega_c,-\omega_1-\omega_c,t_e)
\chi_n(\omega_2-\omega_c,-\omega_2+\omega_c,t_e)\\
+\,&\chi_n(\omega_1+\omega_c,-\omega_2-\omega_c,t_e)
\chi_n(-\omega_2-\omega_c,-\omega_1+\omega_c,t_e)\\
-\,&\chi_n(\omega_1+\omega_c,-\omega_1-\omega_c,t_e)
\chi_n(\omega_2+\omega_c,-\omega_2-\omega_c,t_e)\\
-\,&\chi_n(\omega_1+\omega_c,\omega_2-\omega_c,t_e)
\chi_n(-\omega_1+\omega_c,-\omega_2-\omega_c,t_e)\\
-\,&\chi_n(\omega_1+\omega_c,\omega_2-\omega_c,t_e)
\chi_n(-\omega_2+\omega_c,-\omega_1-\omega_c,t_e),
\big]\big\},
\end{aligned}
\end{equation}
and
\begin{equation}
\begin{aligned}
F_{2,ng}(\vec \omega_3,\omega_c,t_e,2n)\,= \frac{1}{2^6}\big\{
-&\chi_{2n}(\omega_c-\omega_1-\omega_2-\omega_3,-\omega_c+\omega_1,t_e)\,
\chi_{2n}(\omega_c+\omega_2,-\omega_c+\omega_3,t_e)\\
+&\chi_{2n}(\omega_c-\omega_1-\omega_2-\omega_3,-\omega_c+\omega_1,t_e)\,
\chi_{2n}(-\omega_c+\omega_2,\omega_c+\omega_3,t_e)\\
+&\chi_{2n}(\omega_c-\omega_1-\omega_2-\omega_3,\omega_c+\omega_1,t_e)\,
\chi_{2n}(\omega_c+\omega_2,-\omega_c+\omega_3,t_e)\\
-&\chi_{2n}(\omega_c-\omega_1-\omega_2-\omega_3,\omega_c+\omega_1,t_e)\,
\chi_{2n}(-\omega_c+\omega_2,\omega_c+\omega_3,t_e)\big\}.
\end{aligned}
\end{equation}

\bibliography{biblio_finale}
\end{document}